\documentstyle[twocolumn,prl,aps]{revtex}
 
 \newcommand{\BEQ}{\begin{equation}}
 \newcommand{\EEQ}{\end{equation}}
 \newcommand{\BEA}{\begin{eqnarray}}
 \newcommand{\EEA}{\end{eqnarray}}
 \newcommand{\n}{\text{\footnotesize N}}
 \newcommand{\e}{\text{\footnotesize E}}
 \newcommand{\s}{\text{\footnotesize S}}
 \newcommand{\w}{\text{\footnotesize W}}
 \newcommand{\A}{\rm A}
 \newcommand{\B}{\rm B}
 \newcommand{\C}{\rm C}
 \newcommand{\D}{\rm D}
 
\begin{document}
 \draft
 \title{Critical Behavior of the Sandpile Model as a Self-Organized Branching 
 Process}
 \author{E.V. Ivashkevich}
 \address{Laboratory of Theoretical Physics, Joint Institute for Nuclear 
 Research, Dubna, 141980, Russia}
 \date{December 18, 1995}
 \maketitle 
 \begin{abstract}
 Kinetic equations, which explicitly take into account the branching
 nature of sandpile avalanches, are derived. The dynamics of the sandpile
 model is described by the generating functions of a branching process.
 Having used the results obtained the renormalization group approach to the
 critical behavior of the sandpile model is generalized in order to calculate
 both critical exponents and height probabilities.
 \end{abstract}

 \narrowtext
 
 As it was realized a few years ago, almost all fractal structures we can
 see in nature are "self-organized" in a sense that they spontaneously
 grow, obeying rather simple dynamical rules \cite{EPV}. The concept of
 Self-Organized Criticality (SOC) has recently been introduced by Bak
 {\it et al.} \cite{BTW} to emphasize the fact that the production of
 self-similar structures is not the only possible result of the irreversible
 dynamics of an extended system. They have shown that such a system can also
 evolve stochastically into a certain critical state similar to that of second
 order phase transition. It lacks therein any characteristic length as well as
 timescale and obeys power-law distributions. The critical state is
 independent of the initial configuration of the system and, unlike ordinary
 critical phenomena, no fine tuning of control parameters is necessary to
 arrive at this state.
 
 To illustrate the basic ideas of SOC, Bak {\it et al.} used a cellular
 automaton now commonly known as "sandpile" because of the crude analogy
 between its dynamical rules and the way sand topples when building a real
 sand pile. The formulation of this model is given in terms of integer
 height variables $z_i$ at each site of a square lattice ${\cal L}$.
 Particles are added randomly and the addition of a particle increases the
 height at that site by one. If this height exceeds the critical value
 $z_c=4$, then the site topples, and on toppling its height decreases by $4$
 and the heights at each of its nearest neighbors increases by $1$. They may
 become unstable in their turn and the dynamical process continues. Open
 boundary conditions are usually assumed, so that particles can leave the
 system.
 
 Dhar \cite{DD} has shown that this model is exactly solvable due to an
 Abelian group structure hidden in its dynamics. The fact that almost all
 characteristics of the Abelian sandpile model are calculable analytically is
 the reason for considering this model as a perfect proving ground for various
 approaches to SOC. Among others, the Renormalization Group (RG) approach
 proposed recently by Pietronero {\it et al.} \cite{PVZ,LPVZ} seems to be
 the most promising one because it explicitly describes the self-similar
 dynamics of the SOC models at the critical state.
 
 The purpose of this Letter is to generalize this renormalization scheme
 by exploiting the analogy between large scale dynamics of the sandpile
 model and chain reactions of a special kind. At first, we should define
 {\it coarse grained variables} proper for describing the behavior of
 the system on a sublattice ${\cal L}_b$ which consists of the cells of
 size $b$ on the initial lattice ${\cal L}$. To keep the connection
 with the original formulation of the sandpile model, we will characterize
 the static properties of a cell by four quantities
 \BEQ
 {\bf n}^{(b)} = (n_{\A}, n_{\B}, n_{\C}, n_{\D}),~~~
 n_{\A} + n_{\B} + n_{\C} + n_{\D} = 1,
 \EEQ
 which are nothing but the probabilities for a cell to behave
 like a site on the initial lattice with a height 1, 2, 3 or 4
 respectively in the coarse grained dynamics, {\it i.e.} the addition of a 
 "coarse grained
 particle" to the cell transforms it to the next one in the alphabet.
 For example, the cell B characterized by the vector (0,1,0,0) will
 be transformed to the cell C with the vector (0,0,1,0).
 The last variable $n_{\D}$ is the probability for the cell to behave
 like a critical one in a sense that the addition of a "coarse grained 
 particle"
 to the cell induces relaxations into some neighboring cells
 or, in other words, subrelaxation processes on a
 minimal scale span the cell and transfer energy to some of its neighbors.
 
 According to Pietronero {\it et al.} \cite{PVZ},
 independently of the dynamics of the model at the minimal scale such a
 relaxation process leads to four possible situations for a coarse grained
 cell of size $b$. Namely, after the relaxation
 of the critical cell "coarse grained particles" can be transferred to one,
 two, three or four neighboring cells with the probabilities
 \BEQ
 {\bf p}^{(b)} = (p_1, p_2, p_3, p_4),~~~~~  p_1 + p_2 + p_3 + p_4 = 1.
 \label{p}
 \EEQ
 The probabilities $p_i$ are normalized sums of all the
 processes that ignite the corresponding number of neighboring cells,
 independently of their position. The distribution of particles after toppling
 at the minimal scale is characterized by the vector ${\bf p}^{(1)}=(0,0,0,1)$.
 
 While the first set of the coarse grained variables ${\bf n}^{(b)}$ describes
 the height configuration at the scale $b$, the second set ${\bf p}^{(b)}$ 
 defines
 the toppling rules of the model and, in some sense, characterizes the phase
 space for the relaxation dynamics at the same scale.
 
 In this framework the coarse grained dynamics of the sandpile model can be
 represented as the following branching process on the sublattice ${\cal L}_b$
 \BEA
 {\rm A} + \varphi & \rightarrow & {\rm B}, \nonumber\\
 {\rm B} + \varphi & \rightarrow & {\rm C}, \nonumber\\
 {\rm C} + \varphi & \rightarrow & {\rm D}, \\
 {\rm D} + \varphi & \rightarrow &
 \left\
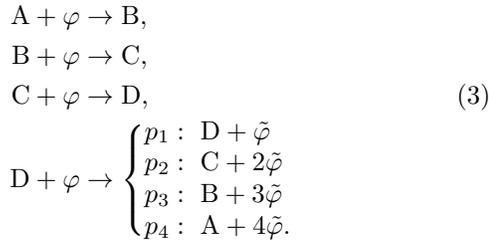
{
 \begin{minipage}{2.3cm}
 $p_1:~ {\rm D} +  \tilde{\varphi}$\\
 $p_2:~ {\rm C} + 2\tilde{\varphi}$\\
 $p_3:~ {\rm B} + 3\tilde{\varphi}$\\
 $p_4:~ {\rm A} + 4\tilde{\varphi}.$
 \end{minipage}
 \right.  \nonumber
 \EEA
 Here $\varphi$ and $\tilde{\varphi}$ denote the "coarse grained particles"
 obtained by the cell and the particles transferred to the neighboring cells,
 respectively.
 
 These processes can formally be reinterpreted as an irreversible chemical
 reaction which takes place at each cell of the sublattice ${\cal L}_b$.
 Now the coarse grained variables $n_{\A}, n_{\B}, n_{\C}, n_{\D}$ and
 $n_\varphi$ denote the concentrations of the respective species
 A, B, C, D, and $\varphi$. Following to standard prescriptions of the
 chemical physics we can write kinetic equations corresponding to this
 scheme of chemical reactions
 \begin{mathletters}
 \BEA
 \dot{n}_{\A} &=& n_\varphi~ (p_4~ n_{\D} - n_{\A}), \label{chem1}\\
 \dot{n}_{\B} &=& n_\varphi~ (p_3~ n_{\D} + n_{\A} - n_{\B}), \label{chem2}\\
 \dot{n}_{\C} &=& n_\varphi~ (p_2~ n_{\D} + n_{\B} - n_{\C}), \label{chem3}\\
 \dot{n}_{\D} &=& n_\varphi~ (p_1~ n_{\D} + n_{\C} - n_{\D}), \label{chem4}\\
 \dot{n}_\varphi &=& n_\varphi~ (\bar{p}~n_{\D} - 1) + \bar{p}~ \nu \nabla^2 
 (n_\varphi n_{\D}) +
 \eta ({\bf r}, t) \label{chem5}
 \EEA
 \end{mathletters}
 where $\bar{p}=p_1+2p_2+3p_3+4p_4$ is equal to the average number of
 particles leaving the cell on toppling and ${\bf r}$ is the position vector
 of the cell in the 2D space. The noise term $\eta({\bf r}, t)$,
 being non-negative, mimics the random addition of particles to the system.
 The diffusion term $\nabla^2 (n_\varphi n_{\D})$ describes the transfer
 of particles into the neighboring cells, and the diffusion coefficient $\nu$
 for the discrete Laplacian on the square lattice is equal to $1/4$.
 
 The only mobile specie in this scheme of reactions is $\varphi$ and it
 is the field $n_\varphi$ which describes the dynamics of avalanches. When it
 is equal to zero, all toppling processes die. Then, due to the noise
 term $\eta({\bf r},t)$, particles are added randomly into the system
 initiating a branching process directed to the open boundary of the
 system. This process mutates species in the cells it has visited and topples
 the critical ones. Finally, the system will reach the steady state where
 the probability that the activity will die is on average balanced by the
 probability that the activity will branch. Thus, the chain reaction maintains
 this stationary state and all further avalanches cannot change the
 concentrations of species A, B, C, and D. Therefore, the steady
 state is characterized by the conditions that
 \BEQ
 \dot{n}_{\A}=\dot{n}_{\B}=\dot{n}_{\C}=\dot{n}_{\D}=0
 \EEQ
 and Eqs.\ (\ref{chem1}-\ref{chem4}) lead to the following relationships
 between concentrations of species ${\bf n}^{(b)}$ at the stationary state and
 branching probabilities ${\bf p}^{(b)}$
 \begin{mathletters}
 \BEA
 n_{\A}^* &=& p_4/\bar{p},  \label{balanceA}\\
 n_{\B}^* &=& (p_3+p_4)/\bar{p},  \label{balanceB}\\
 n_{\C}^* &=& (p_2+p_3+p_4)/\bar{p}, \label{balanceC}\\
 n_{\D}^* &=& (p_1+p_2+p_3+p_4)/\bar{p}=1/\bar{p}~. \label{balanceD}
 \EEA
 \end{mathletters}
 The relation (\ref{balanceD}) between the probability $n_{\D}^*$
 and branching probabilities
 ${\bf p}^{(b)}$ has already appeared in the paper \cite{PVZ}. It was
 derived there from the assumption that at the stationary state the
 flow of particles in a cell was on average balanced by the flow of
 particles out of the cell.
 
 If we neglect the fluctuations of ${\bf n}^{(b)}$ at the steady state,
 Eq.\ (\ref{chem5}), describing the propagation of an active process,
 becomes simply the diffusion equation
 \BEQ
 \dot{n}_\varphi =  \nu \nabla^2 (n_\varphi) + \eta ({\bf r}, t)
 \EEQ
 and coincides with that for the flow of particles at the critical state
 obtained by Zhang \cite{YCZ}.
 
 To describe in detail the branching process underlying the large scale
 behavior of the sandpile model, let us consider the following generating
 function:
 \BEA
 \lefteqn{\sigma(\n,\e,\s,\w)=\frac{p_1}{4}(\n+\e+\s+\w)}\nonumber\\
 & & ~~~~+\frac{p_2}{6}(\n\e+\n\s+\n\w+\e\s+\e\w+\s\w) \label{gf}\\
 & & ~~~~+\frac{p_3}{4}(\n\e\s+\n\e\w+\n\s\w+\e\s\w)+p_4~\n\e\s\w,
 \nonumber
 \EEA
 where symbols $\n,\e,\s$ and $\w$ correspond to the north, east, south
 and west directions on the square lattice, respectively. The coefficient
 of each term of this polynomial, say $p_2\n\e/6$, is equal to the
 probability that after relaxation of a critical cell particles will
 go only northward and eastward.
 
 It is easy to check directly that this function has the following
 properties: (a) if any letter, for example $\n$, is replaced by zero,
 the function $\sigma(0,\e,\s,\w)$ counts all the processes that do not
 send a particle northward; (b) on the contrary, the difference
 $\sigma(1,\e,\s,\w)-\sigma(0,\e,\s,\w)$ corresponds to the sum over all
 the processes that definitely send a particle to the north and,
 possibly, to some other directions; (c) this generating function is
 normalized so that $\sigma(1,1,1,1)=1$ due to the normalization condition
 on branching probabilities, Eq.\ (\ref{p}); (d) if the particle that has
 been sent northward after the relaxation of a critical cell ignites the
 northern neighboring cell, the corresponding letter $\n$ should be
 replaced by another $\sigma$-function and each term of the generating
 function $\sigma(\sigma(\n,\e,\s,\w),\e,\s,\w)-\sigma(0,\e,\s,\w)$
 corresponds to the process that consists of the two successive
 topplings.
 
 Using these simple rules we can write the generating function for any
 chain of relaxation processes on the sublattice ${\cal L}_b$.
 
 Now, following to the general ideas of the paper \cite{PVZ}, we are
 ready to define a renormalization transformation for the relaxation
 dynamics of the sandpile model. The standard real space RG approach
 consists of considering a block $2\times2$ of cells of the lattice
 ${\cal L}_{b}$ to be a single cell of the size $2b$ on the lattice
 ${\cal L}_{2b}$. Thus, given branching probabilities ${\bf p}^{(b)}$
 we want to find an analogous set of probabilities ${\bf p}^{(2b)}$ on
 the lattice ${\cal L}_{2b}$.  To this end, we should count up all the
 possible toppling processes that span the starting block of cells and
 transfer particles to some neighboring blocks. The spanning rule
 implies that we have to consider only those connected chains of
 processes that span the block from left to right or from top to bottom
 neglecting the processes not extending over the resulting scale $2b$.
 On Fig.\ \ref{fig} all the different types of toppling processes that
 satisfy this spanning condition are shown. The blocks that have no, or
 have only one critical cell inside are not included in the
 renormalization of the dynamics because they do not lead to a
 relaxation process that spans the cell of size $2b$.
 
 We amplify the set of toppling processes considered in the RG scheme of
 Pietronero {\it et al.} \cite{PVZ} by taking into account also the
 processes, Fig.\ \ref{fig} (c), of relaxation of the cell C provided
 it has got two particles during the relaxation of the block. These additional
 processes seem to be very important for the relaxation dynamics of the
 sandpile model, and taking account of them improves the accuracy of the
 RG approach considerably.
 
 \begin{figure*}
 \unitlength=0.75mm
 \special{em:linewidth 0.5pt}
 \linethickness{0.5pt}
 \begin{picture}(106.00,120.00)
 \thinlines
 \put(14.00,98.00){\framebox(20.00,20.00)[cc]{}}
 \put(19.00,113.00){\circle{5.66}}
 \put(29.00,113.00){\circle{5.66}}
 \thicklines
 \put(19.00,103.00){\circle{5.66}}
 \thinlines
 \put(29.00,103.00){\circle{5.66}}
 \put(19.00,113.00){\makebox(0,0)[cc]{\scriptsize D}}
 \put(19.00,103.00){\makebox(0,0)[cc]{\scriptsize \bf D}}
 \put(29.00,103.00){\makebox(0,0)[cc]{\scriptsize X}}
 \put(29.00,113.00){\makebox(0,0)[cc]{\scriptsize X}}
 \put(19.00,106.00){\vector(0,1){4.00}}
 \put(14.00,67.00){\framebox(20.00,20.00)[cc]{}}
 \put(19.00,82.00){\circle{5.66}}
 \put(29.00,82.00){\circle{5.66}}
 \thicklines
 \put(19.00,72.00){\circle{5.66}}
 \thinlines
 \put(29.00,72.00){\circle{5.66}}
 \put(19.00,82.00){\makebox(0,0)[cc]{\scriptsize D}}
 \put(19.00,72.00){\makebox(0,0)[cc]{\scriptsize \bf D}}
 \put(29.00,72.00){\makebox(0,0)[cc]{\scriptsize X}}
 \put(29.00,82.00){\makebox(0,0)[cc]{\scriptsize D}}
 \put(19.00,75.00){\vector(0,1){4.00}}
 \put(49.00,67.00){\framebox(20.00,20.00)[cc]{}}
 \put(54.00,82.00){\circle{5.66}}
 \put(64.00,82.00){\circle{5.66}}
 \thicklines
 \put(54.00,72.00){\circle{5.66}}
 \thinlines
 \put(64.00,72.00){\circle{5.66}}
 \put(54.00,82.00){\makebox(0,0)[cc]{\scriptsize D}}
 \put(54.00,72.00){\makebox(0,0)[cc]{\scriptsize \bf D}}
 \put(64.00,72.00){\makebox(0,0)[cc]{\scriptsize D}}
 \put(64.00,82.00){\makebox(0,0)[cc]{\scriptsize X}}
 \put(54.00,75.00){\vector(0,1){4.00}}
 \put(14.00,36.00){\framebox(20.00,20.00)[cc]{}}
 \put(19.00,51.00){\circle{5.66}}
 \put(29.00,51.00){\circle{5.66}}
 \thicklines
 \put(19.00,41.00){\circle{5.66}}
 \thinlines
 \put(29.00,41.00){\circle{5.66}}
 \put(19.00,51.00){\makebox(0,0)[cc]{\scriptsize D}}
 \put(19.00,41.00){\makebox(0,0)[cc]{\scriptsize \bf D}}
 \put(29.00,41.00){\makebox(0,0)[cc]{\scriptsize C}}
 \put(29.00,51.00){\makebox(0,0)[cc]{\scriptsize D}}
 \put(19.00,44.00){\vector(0,1){4.00}}
 \put(49.00,36.00){\framebox(20.00,20.00)[cc]{}}
 \put(54.00,51.00){\circle{5.66}}
 \put(64.00,51.00){\circle{5.66}}
 \thicklines
 \put(54.00,41.00){\circle{5.66}}
 \thinlines
 \put(64.00,41.00){\circle{5.66}}
 \put(54.00,51.00){\makebox(0,0)[cc]{\scriptsize D}}
 \put(54.00,41.00){\makebox(0,0)[cc]{\scriptsize \bf D}}
 \put(64.00,41.00){\makebox(0,0)[cc]{\scriptsize D}}
 \put(64.00,51.00){\makebox(0,0)[cc]{\scriptsize C}}
 \put(54.00,44.00){\vector(0,1){4.00}}
 \put(14.00,5.00){\framebox(20.00,20.00)[cc]{}}
 \put(19.00,20.00){\circle{5.66}}
 \put(29.00,20.00){\circle{5.66}}
 \thicklines
 \put(19.00,10.00){\circle{5.66}}
 \thinlines
 \put(29.00,10.00){\circle{5.66}}
 \put(19.00,20.00){\makebox(0,0)[cc]{\scriptsize D}}
 \put(19.00,10.00){\makebox(0,0)[cc]{\scriptsize \bf D}}
 \put(29.00,10.00){\makebox(0,0)[cc]{\scriptsize D}}
 \put(29.00,20.00){\makebox(0,0)[cc]{\scriptsize D}}
 \put(19.00,13.00){\vector(0,1){4.00}}
 \put(22.00,82.00){\vector(1,0){4.00}}
 \put(57.00,72.00){\vector(1,0){4.00}}
 \put(22.00,51.00){\vector(1,0){4.00}}
 \put(57.00,41.00){\vector(1,0){4.00}}
 \put(29.00,48.00){\vector(0,-1){4.00}}
 \put(22.00,41.00){\vector(1,0){4.00}}
 \put(57.00,51.00){\vector(1,0){4.00}}
 \put(64.00,44.00){\vector(0,1){4.00}}
 \put(22.00,20.00){\vector(1,0){4.00}}
 \put(24.00,94.00){\makebox(0,0)[cc]{\scriptsize X=A,B,C}}
 \put(24.00,63.00){\makebox(0,0)[cc]{\scriptsize X=A,B}}
 \put(59.00,63.00){\makebox(0,0)[cc]{\scriptsize X=A,B}}
 \put(84.00,67.00){\framebox(20.00,20.00)[cc]{}}
 \put(89.00,82.00){\circle{5.66}}
 \put(99.00,82.00){\circle{5.66}}
 \thicklines
 \put(89.00,72.00){\circle{5.66}}
 \thinlines
 \put(99.00,72.00){\circle{5.66}}
 \put(89.00,72.00){\makebox(0,0)[cc]{\scriptsize \bf D}}
 \put(99.00,82.00){\makebox(0,0)[cc]{\scriptsize D}}
 \put(94.00,63.00){\makebox(0,0)[cc]{\scriptsize X=A,B}}
 \put(84.00,36.00){\framebox(20.00,20.00)[cc]{}}
 \put(89.00,51.00){\circle{5.66}}
 \put(99.00,51.00){\circle{5.66}}
 \thicklines
 \put(89.00,41.00){\circle{5.66}}
 \thinlines
 \put(99.00,41.00){\circle{5.66}}
 \put(89.00,41.00){\makebox(0,0)[cc]{\scriptsize \bf D}}
 \put(99.00,51.00){\makebox(0,0)[cc]{\scriptsize D}}
 \put(89.00,44.00){\vector(0,1){4.00}}
 \put(92.00,41.00){\vector(1,0){4.00}}
 \put(89.00,82.00){\makebox(0,0)[cc]{\scriptsize X}}
 \put(99.00,72.00){\makebox(0,0)[cc]{\scriptsize D}}
 \put(92.00,72.00){\vector(1,0){4.00}}
 \put(99.00,75.00){\vector(0,1){4.00}}
 \put(89.00,51.00){\makebox(0,0)[cc]{\scriptsize C}}
 \put(99.00,41.00){\makebox(0,0)[cc]{\scriptsize D}}
 \put(99.00,44.00){\vector(0,1){4.00}}
 \put(96.00,51.00){\vector(-1,0){4.00}}
 \put(49.00,5.00){\framebox(20.00,20.00)[cc]{}}
 \put(54.00,20.00){\circle{5.66}}
 \put(64.00,20.00){\circle{5.66}}
 \thicklines
 \put(54.00,10.00){\circle{5.66}}
 \thinlines
 \put(64.00,10.00){\circle{5.66}}
 \put(54.00,20.00){\makebox(0,0)[cc]{\scriptsize D}}
 \put(54.00,10.00){\makebox(0,0)[cc]{\scriptsize \bf D}}
 \put(64.00,10.00){\makebox(0,0)[cc]{\scriptsize D}}
 \put(64.00,20.00){\makebox(0,0)[cc]{\scriptsize D}}
 \put(54.00,13.00){\vector(0,1){4.00}}
 \put(57.00,20.00){\vector(1,0){4.00}}
 \put(84.00,5.00){\framebox(20.00,20.00)[cc]{}}
 \put(89.00,20.00){\circle{5.66}}
 \put(99.00,20.00){\circle{5.66}}
 \thicklines
 \put(89.00,10.00){\circle{5.66}}
 \thinlines
 \put(99.00,10.00){\circle{5.66}}
 \put(89.00,20.00){\makebox(0,0)[cc]{\scriptsize D}}
 \put(89.00,10.00){\makebox(0,0)[cc]{\scriptsize \bf D}}
 \put(99.00,10.00){\makebox(0,0)[cc]{\scriptsize D}}
 \put(99.00,20.00){\makebox(0,0)[cc]{\scriptsize D}}
 \put(29.00,17.00){\vector(0,-1){4.00}}
 \put(57.00,10.00){\vector(1,0){4.00}}
 \put(64.00,13.00){\vector(0,1){4.00}}
 \put(92.00,10.00){\vector(1,0){4.00}}
 \put(99.00,13.00){\vector(0,1){4.00}}
 \put(96.00,20.00){\vector(-1,0){4.00}}
 \put(12.00,3.00){\dashbox{1.00}(94.00,24.00)[cc]{}}
 \put(0.00,108.00){\makebox(0,0)[lc]{(a)}}
 \put(0.00,77.00){\makebox(0,0)[lc]{(b)}}
 \put(0.00,46.00){\makebox(0,0)[lc]{(c)}}
 \put(0.00,15.00){\makebox(0,0)[lc]{(d)}}
 \end{picture}
 \caption{We show the four different types of the blocks and some relaxation
 schemes spanning them. The other schemes can be obtained from these figures
 by rotations. It is convenient for calculations to subdivide the relaxation
 processes in the block (d) into three parts shown in the dashed box.}
 \label{fig}
 \end{figure*}
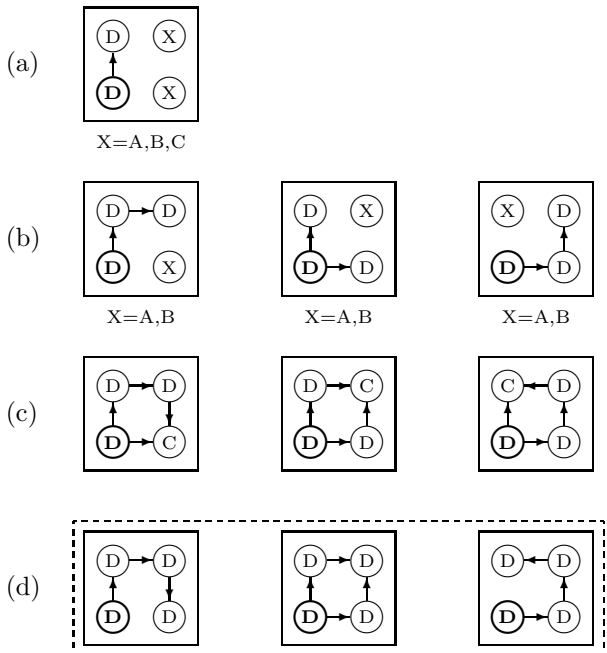
 
 At the stationary state each type of blocks has an additional
 weight given by the product of the probabilities ${\bf n}^{(b)}$ of the
 cells inside the block and the numerical factor equal to the numbers of
 different blocks with the same relaxation schemes. Thus, to the blocks on
 Fig.\ \ref{fig} the following relative weights should be ascribed
 \begin{mathletters}
 \BEA
 W_a&=&4 n_{\D}^2 (n_{\A}+n_{\B}+n_{\C})^2,\label{weight1}\\
 W_b&=&4 n_{\D}^3 (n_{\A}+n_{\B}), \label{weight2}\\
 W_c&=&4 n_{\D}^3 n_{\C}, \label{weight3}\\
 W_d&=&  n_{\D}^4.\label{weight4}
 \EEA
 \end{mathletters}
 The generating function corresponding to the relaxation processes inside
 the blocks with the weight $W_a$ can be written with the use of the
 $\sigma$-function, Eq.\ (\ref{gf}),
 \BEA
 \lefteqn{\Sigma_a(\n,\e,\s,\w)=} \nonumber \\
 & &~ \{\sigma(\sigma(\n,1,1,\w),1,\s,\w)-\sigma(0,1,\s,\w)+{c.p.}\}/N_a,
 \EEA
 where the notation $c.p.$ implies all possible cyclic permutations of the
 symbols $\n, \e, \s$ and $\w$ and $N_a$ is the normalization factor such
 that $\Sigma_a(1,1,1,1)=1$. To write this function we start from the left
 down cell and define the arguments of the $\sigma$-function corresponding
 to the toppling of this cell. For the process definitely spans the block,
 the left up cell should topple and we write another $\sigma$-function
 instead of the symbol $\n$. By going eastward the process will terminate
 inside the block and this branch of the toppling process cannot affect
 the neighboring blocks. Hence, we should write the number $1$ instead of
 the symbol $\e$. The other symbols $\s$ and $\w$ correspond to the branches
 of the toppling process which immediately get out the initial block of cells.
 
 Analogously, we can write the generating function of the blocks with the
 weight $W_b$
 \BEA
 \lefteqn{\Sigma_b(\n,\e,\s,\w)=} \nonumber \\
 & &~ \{\sigma(\sigma(\n,\sigma(\n,\e,1,1),1,\w),1,\s,\w)-\sigma(0,1,\s,\w)
\nonumber \\
 & &~ +\sigma(\sigma(\n,1,1,\w),\sigma(1,\e,\s,1),\s,\w)-\sigma(0,0,\s,\w)  \\
 & &~ +\sigma(1,\sigma(\sigma(\n,\e,1,1),\e,\s,1),\s,\w)-\sigma(1,0,\s,\w) 
 \nonumber\\
 & &~ +{c.p.}\}/N_b. \nonumber
 \EEA
 The generating functions $\Sigma_c$ and $\Sigma_d$ for the blocks with the
 weights $W_c$ and $W_d$ in Fig.\ \ref{fig} are quite similar in principle
 but the expressions are rather long and will be published in a complete paper.
 
 All these generating functions are polynomials describing in detail
 toppling processes inside the blocks. As it is easy to check directly,
 besides the terms with the first powers of the symbols $\n, \e, \s$ and $\w$
 these generating functions have also the terms with their squares. The first
 terms correspond to the processes when after the relaxation of the block only 
 one
 particle goes in a given direction, while the last ones describe the processes
 with two particles going in the same direction. According to the RG ideology,
 we should consider both of them as the transfer of the only new
 "coarse grained particle". To this end, we can simply replace all the squares
 of the symbols $(\n, \e, \s, \w)$ by their first powers. Finally, we will get
 polynomials of the same form as the original one, Eq.\ (\ref{gf}), but with
 the new branching probabilities.
 
 Now, to write the complete generating function corresponding to the
 relaxation of the block $2\times 2$ of the cells on the lattice ${\cal L}_b$,
 we should average the $\Sigma$-functions with the corresponding weights of
 blocks defined by the height probabilities ${\bf n}^{(b)}$ at the scale
 $b$, Eqs.\ (\ref{weight1}-\ref{weight4}),
 \BEA
 \lefteqn{
 \Sigma(\n,\e,\s,\w)=\{W_a\Sigma_a(\n,\e,\s,\w)+W_b\Sigma_b(\n,\e,\s,\w)}
 \nonumber\\
 & &~~~+W_c\Sigma_c(\n,\e,\s,\w)+W_d\Sigma_d(\n,\e,\s,\w)\}/N.
 \EEA
 
 To finish the whole RG scheme we have to define an analogous renormalization
 equations for the height probabilities ${\bf n}^{(2b)}$.
 Due to the nonlocal properties of the dynamics of the sandpile model, the
 direct renormalization of these quantities seems to be very difficult.
 Instead, we will use the stationary state relations, Eqs.\
 (\ref{balanceA}-\ref{balanceD}), to define the renormalized height
 probabilities.
 
 Given this RG transformation we can study how the system evolve under the
 successive doubling of the scale. The final result is that independent of
 the vectors ${\bf p}^{(1)}$ and ${\bf n}^{(1)}$ at the minimal scale the
 system flows at the same nontrivial fixed point. This fixed point is
 attractive in the whole phase space and the system evolves spontaneously
 toward the critical values ${\bf p}^*$ and ${\bf n}^*$ shown in
 Table~\ref{table}.\\
 \parbox{7.5cm}{
 \begin{table}
 \caption{Fixed point parameters of the RG transformation in comparison
 with analogous exact results.}
 \begin{tabular}{lcccc}
       & $p_1^*$ & $p_2^*$ & $p_3^*$ & $p_4^*$  \\
 \hline\\[-0.22cm]
 RG               & 0.295 & 0.435 & 0.229 & 0.041  \\
 exact \cite{MDM} & 0.295 & 0.447 & 0.222 & 0.036\\[0.1cm]
       &$n_{\A}^*$&$n_{\B}^*$&$n_{\C}^*$&$n_{\D}^*$\\
 \hline\\[-0.22cm]
 RG               & 0.021 & 0.134 & 0.349 & 0.496  \\
 exact \cite{VBP} & 0.074 & 0.174 & 0.306 & 0.446\\
 \end{tabular}
 \label{table}
 \end{table}
 }\\
 These results obtained with the use of the RG approach can be compared 
 with the
 exact ones for the sandpile model. As it has been shown in the paper 
 \cite{IKP},
  an avalanche can be considered as a sequence of waves of topplings each
 consisting of sites that toppled only once in that wave. Being more simple
 objects, waves admit a representation in terms of spanning trees covering
 the lattice sites. The branching probabilities of the spanning trees have
 been calculated exactly by Manna {\it et al.} \cite{MDM}. Their results are
 presented in Table~\ref{table}. The hypothesis that branching
 probabilities for spanning trees should coincide with that for the toppling
 process seems to be quite plausible, but it has yet to be proved. The exact
 height probabilities for the sandpile model presented in Table
 \ref{table} were calculated by Priezzhev \cite{VBP}.
 
 Recently, Priezzhev {\it et al.} \cite{PKI} have used the known exponents
 for spanning trees to argue that in the sandpile model the probability
 for the number of sites to involve in the avalanche equal to $s$ varies
 as $P(s)\approx s^{-\tau}$, for large $s$, where $\tau=5/4$.
 
 As it has been noted by Pietronero {\it et al.} \cite{PVZ} the avalanche 
 exponent
 $\tau$ can be obtained directly from the fixed point parameters. Below, we
 briefly repeat their arguments.
 By using the discrete length scale $b^{(k)}=2^k$ and the
 avalanche distribution in the form $P(r)dr\approx r^{(1-2\tau)}dr$ we can
 define the probability that the relaxation process spans the cell of size
 $b^{(k)}$ and dies at the neighboring cells not extending over the
 scale $b^{(k+1)}$
 \BEQ
 K=\int_{b^{(k-1)}}^{b^{(k)}}P(r)dr/\int_{b^{(k-1)}}^{\infty}P(r)dr
 =1-2^{2(1-\tau)}
 \label{K1}
 \EEQ
 Asymptotically $(k\rightarrow\infty)$ we can express $K$ in terms of fixed
 point
 parameters in the following way:
 \BEA
 \lefteqn{K=p_1^*(1-n_{\D}^*)+p_2^*(1-n_{\D}^*)^2}\nonumber\\
 & &~~~+p_3^*(1-n_{\D}^*)^3+p_4^*(1-n_{\D}^*)^4.
 \label{K2}
 \EEA
 Using these two expressions, Eqs.\ (\ref{K1},\ref{K2}), the exponent 
 $\tau$ is
 given by the formula
 \BEQ
 \tau=1-\frac{1}{2}\frac{\ln(1-K)}{\ln2}=1.248
 \EEQ
 in excellent agreement with the proposed value $\tau=5/4$.

 It is a pleasure to thank V.B.~Priezzhev, D.V.~Ktitarev, A.~Vespignani and
 V.~Loreto for fruitful discussions.
 This work was supported by the Russian Foundation for Basic Research
 through Grant No. 95-01-0257 and by the International Center for Fundamental
 Physics in Moscow through INTAS Grant No. 93-2492.

 \end{document}